\documentclass[twocolumn]{aastex7}

\begin{document}

\title{A Pilot Study to Verify the RR Lyrae Candidates with Vera C. Rubin Observatory Early Alerts}

\author[0000-0001-8771-7554,gname=Chow-Choong, sname=Ngeow]{Chow-Choong Ngeow}
\affil{Graduate Institute of Astronomy, National Central University, 300 Jhongda Road, 32001 Jhongli, Taiwan}
\affil{Taiwan Astronomical Research Alliance (TARA)}
\email[show]{cngeow@astro.ncu.edu.tw}

\author[0000-0001-6147-3360,gname=Anupam, sname=Bhardwaj]{Anupam Bhardwaj}
\affil{Inter-University Center for Astronomy and Astrophysics (IUCAA), Post Bag 4, Ganeshkhind, Pune 411 007, India}
\email{anupam.bhardwaj@iucaa.in}

\author[0000-0003-1959-8439,gname=Sarang, sname=Shah]{Sarang Shah}
\affil{Inter-University Center for Astronomy and Astrophysics (IUCAA), Post Bag 4, Ganeshkhind, Pune 411 007, India}
\email{sarang.shah@iucaa.in}

\author[0000-0003-4799-5079,gname=Steven, sname=Gough-Kelly]{Steven Gough-Kelly}
\affil{Jeremiah Horrocks Institute, University of Lancashire, Preston PR1 2HE, UK}
\email{sgoughkelly@gmail.com}

\author[0000-0002-3045-0446,gname=Oleksandra, sname=Razim]{Oleksandra Razim}
\affil{Center for Astrophysics and Cosmology, University of Nova Gorica, Vipavska 11c, 5270 Ajdovšcina, Slovenia}
\email{shr.razim@gmail.com}

%\correspondingauthor{Chow-Choong Ngeow}

\begin{abstract}

We present a pilot study using the Vera C. Rubin Observatory early alerts to verify RR Lyrae candidates in the Pan-STARRS1, the Dark Energy Survey, and the Next-Generation Virgo Cluster Survey RR Lyrae catalogs. RR Lyrae candidates fainter than 16~mag in the $g$-band in these catalogs were crossmatched with the alerts observed in several deep drilling fields and the M49 field. After excluding alerts with a low number of detections, there are 40 alerts associated with the RR Lyrae candidates. The multiband Rubin-LSST light curves extracted from the alerts verify 32 variables as genuine RR Lyrae, although several were not classified as variable stars in the {\tt ALeRCE} and {\tt Lasair} community alert brokers. While {\tt ALeRCE} and {\tt Lasair} provide $\sim70\%$ and $\sim40\%$ true variable classification, respectively, we find that $20\%$ of the alert sample are non-RR Lyrae variables. The remaining eight candidate variables do not show typical RR Lyrae light curves and include two active galactic nuclei and two eclipsing binaries. Additionally, we have also found a small number of known variable candidates with no alerts, which would suggest that they are either not RR Lyrae variables or the template images are not yet available for their difference image analysis.
  
\end{abstract}

\section{Introduction}
 
Population II pulsating stars such as RR Lyrae (RRL) are excellent distance indicators \citep[for reviews, see][]{beaton2018,bhardwaj2020}, which have been used to trace the old components of the Milky Way (MW) and investigate the sub-structures of the Galactic halo \citep[for recent examples, see][]{hernitschek2018,iorio2021,navarro2021,stringer2021,wang2022,chen2023,feng2024,medina2024,lucey2026}. Most of these studies relied on distant RRL samples compiled from various synoptic time-domain sky surveys. Therefore, the NSF-DOE Vera C. Rubin Observatory Legacy Survey of Space and Time \citep[hereafter Rubin-LSST,][]{ivezic2019} will revolutionize our understanding of the MW and Local Group galaxies by discovering distant RRL populations up to $\sim600$~kpc \citep{oluseyi2012}.

Most synoptic time-domain surveys provide a limited temporal baseline, and their variable star light curves are either sparsely sampled or exhibit large photometric errors, or both, especially for the distant RRL near the detection limits. As a result, classifications of these faint RRL are prone to contamination from active galactic nuclei (AGN, i.e., AGN could be misidentified as RRL) or other variable sources. This is illustrated in the work of \citet{feng2026}, who studied a sample of 65 distant MW halo RRL using spectroscopic observations. They found that 10 of their RRL candidates turned out to be quasars or blazars, although they had only selected RRL targets with high probability or high scores in various public catalogs, such as the Pan-STARRS1 (PS1) RRL catalog \citep{sesar2017}, the Dark Energy Survey (DES) RRL catalog \citep{stringer2019}, and the Next Generation Virgo Cluster Survey (NGVS) RRL catalog \citep{feng2024}.      

\begin{deluxetable*}{lllcccccc}
  %\movetableright=-1in
  \label{tab1}
  \tabletypesize{\scriptsize}
  \tablecaption{Summary of the Fields and the Selection of Final Alerts in each Fields}
  \tablewidth{10pt}
  \tablehead{
    \colhead{Field} &
    \colhead{$RA$ (deg.)} &
    \colhead{$DEC$ (deg.)} &
    \colhead{RRL Catalog\tablenotemark{a}} &
    \colhead{$N_{RRL}$} &
    \colhead{$\langle g \rangle > 16$} &
    \colhead{$1 < N_{\mathrm{det}} < 50$}&
    \colhead{No alert}&
    \colhead{$N_{\mathrm{Final}}$} \\
    \colhead{(1)} &
    \colhead{(2)} &
    \colhead{(3)} &
    \colhead{(4)} &
    \colhead{(5)} &
    \colhead{(6)} &
    \colhead{(7)} &
    \colhead{(8)} &
    \colhead{(9)} 
  }
  \startdata
  ELAISS1 & 9.45  & $-$44.02 & DES & 9  & 7  & 7 & 0  & 0 \\
  %XMM\_LSS& 35.57 & $-$4.82  & DES,PS1\\
  ECDFS   & 52.98 & $-$28.12 & DES,PS1 & 11 & 5 & 2 & 3 & 0\\
  COSMOS  &150.11 & $+$2.23  & PS1     & 40 & 27& 9 & 6 & 12\\
  EDFS\_a & 58.9  & $-$49.32 & DES     & 9  & 9 & 3 & 1 & 5 \\
  EDFS\_b & 63.6  & $-$47.6  & DES     & 7  & 6 & 1 & 2 & 3 \\
  M49     &187.25 & $+$8.0   & PS1,NGVS& 55 &49 &26 & 3 & 20\\ 
  \enddata 
  \tablenotetext{a}{DES -- \citet{stringer2021}; NGVS -- \citet{feng2024}; PS1 -- \citet{sesar2017}.}
\end{deluxetable*}

\begin{deluxetable*}{lrlcccccc}
  %\movetableright=-1in
  \label{tab_noalert}
  \tabletypesize{\scriptsize}
  \tablecaption{RR Lyrae Candidates without Alerts.}
  \tablewidth{10pt}
  \tablehead{
    \colhead{Field} &
    \colhead{$RA$ (deg.)} &
    \colhead{$DEC$ (deg.)} &
    \colhead{$\Delta \mathrm{R}$ (deg.)\tablenotemark{a}} &
    \colhead{Catalog/ID} & 
    \colhead{Scores\tablenotemark{b}} &
    \colhead{$\langle g \rangle$} &
    \colhead{$\langle r \rangle$} &
    \colhead{$\langle i \rangle$} 
  }
  \startdata
  ECDFS & 53.32123  &-29.05875&	0.985 & PS1/73120533210819793 & 0.86/0.01  &	20.51 & 20.37 & 20.32 \\
  ECDFS & 53.11722  &-29.45599&	1.341 & DES/1396115307 & 0.938/1.000/0.672 &	22.16 & 21.92 & 21.95 \\
  COSMOS & 150.49770 & +2.63516 & 0.561 & PS1/111161504976072626 & 0.42/0.08 & 20.87 & 20.72 & 20.62 \\
  COSMOS & 150.94836 & +2.67063 & 0.946 & PS1/111201509484485113 & 0.43/0.01 & 21.08 & 20.81 & 20.75 \\
  COSMOS & 150.91483 & +2.99884 & 1.112 & PS1/111591509148709076 & 0.50/0.01 & 20.83 & 20.61 & 20.53 \\
  COSMOS & 150.26915 & +0.74510 & 1.493 & PS1/108891502691414581 & 0.30/0.11 & 20.39 & 20.17 & 20.11 \\
  COSMOS & 150.16513 & +3.78673 & 1.558 & PS1/112541501651324509 & 0.01/0.01 & 21.14 & 20.87 & 20.79 \\
  EDFS\_a& 56.49932  &-49.72656& 1.610 & DES/1418341148 & 0.855/0.944/0.732 & 23.11 & 22.96 & 22.74 \\
  M49   & 187.48327 & +9.33104 & 1.351 & NGVS/J122955.99+091951.7 & $\cdots$ & 22.35 & $\cdots$ & 23.31 \\
  \enddata 
  \tablenotetext{a}{Radial distance, in degrees, from the center of a given field as listed in Table \ref{tab1}.}
  \tablenotetext{b}{Scores in the PS1 RRL catalog are listed as $S3ab/S3c$, where $S3ab$ and $S3c$ are the scores for the ab-type and c-type RRL, respectively. Scores in the DES RRL catalog are listed as RF1/RF2/RF3, where RF is the random-forest classification scores. Further details are given in \citet{sesar2017} and \citet{stringer2021}. In the case of the NGVS RRL catalog, \citet{feng2024} only selected RRL candidates with scores larger than 0.93, hence there is no score given in this catalog. Note that a score of one implies the candidate is confidently an RRL.}
\end{deluxetable*}

In addition to using spectroscopic observations, another method to separate AGN from faint RRL candidates is to use deeper, more frequent photometric observations from upcoming time-domain sky surveys. For example, Rubin-LSST will provide a depth of $r\sim24.7$~mag \citep{bianco2022} on single-epoch observations and accumulate a few hundreds of data-points after 10 years of operation. Such a combination is ideal for photometrically separating genuine RRL and AGN for faint variable candidates in the southern sky, because RRL exhibit well-characterized periodic light curves \citep[e.g. Fig. 2 in][]{2022Univ....8..122B}.

Rubin-LSST did not begin its official operation at the time of the preparation of this manuscript. On the other hand, Rubin Observatory started producing alerts \citep{juric2023,guy2025} to the public on 24 February 2026 in several Deep Drilling Fields (DDFs, with the exception of the XMM\_LSS field).\footnote{See \url{https://survey-strategy.lsst.io/baseline/ddf.html}.\label{fn_ddf}} Therefore, we initiated a pilot study to verify the nature of faint RRL candidates in the public RRL catalogs using the light curves extracted from Rubin alerts. Section \ref{sec_sample} presents the RRL candidates and the associated alerts analyzed in this study. Our analysis and results are given in Section \ref{sec_analy}, followed by discussions and conclusions of our study in Section \ref{sec_last}.

\section{Sample of RRL Candidates and Alerts} \label{sec_sample}

\begin{figure*}
  \epsscale{1.0}
  \plottwo{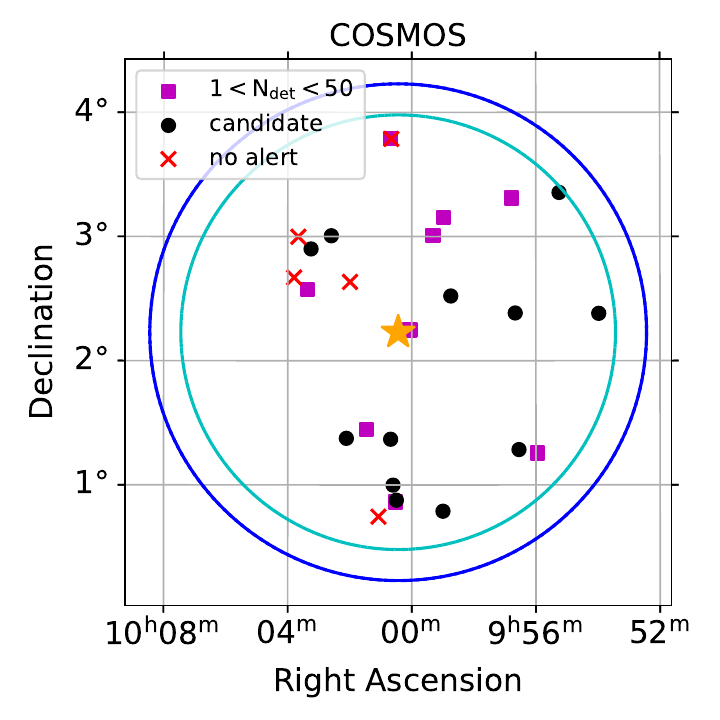}{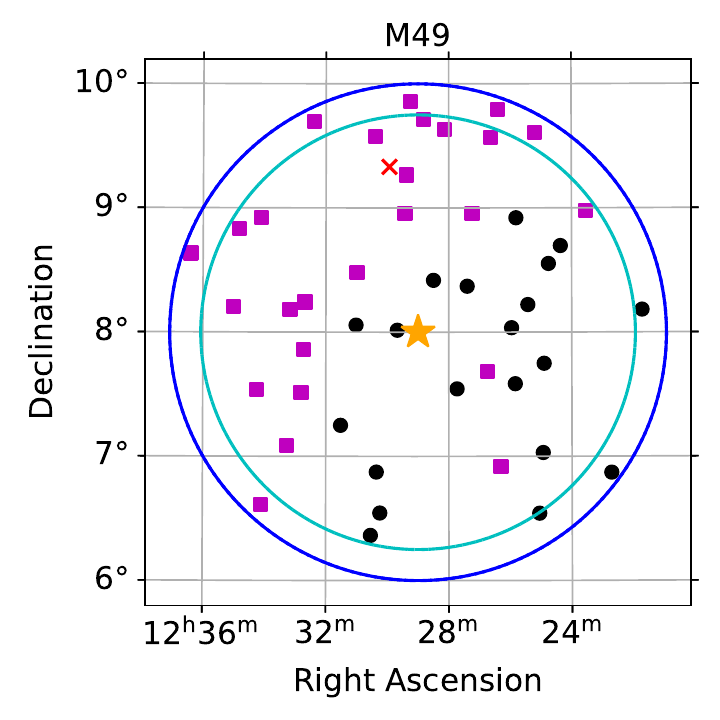}
  \plottwo{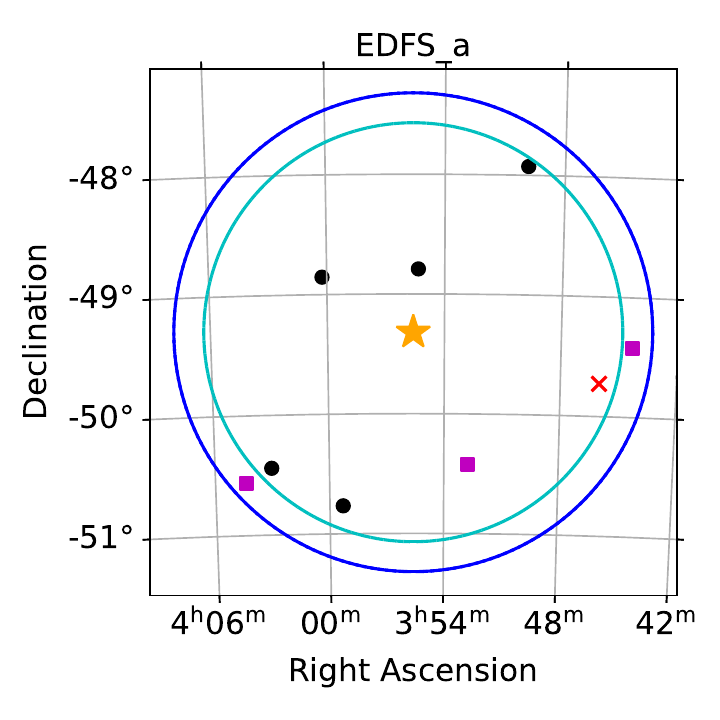}{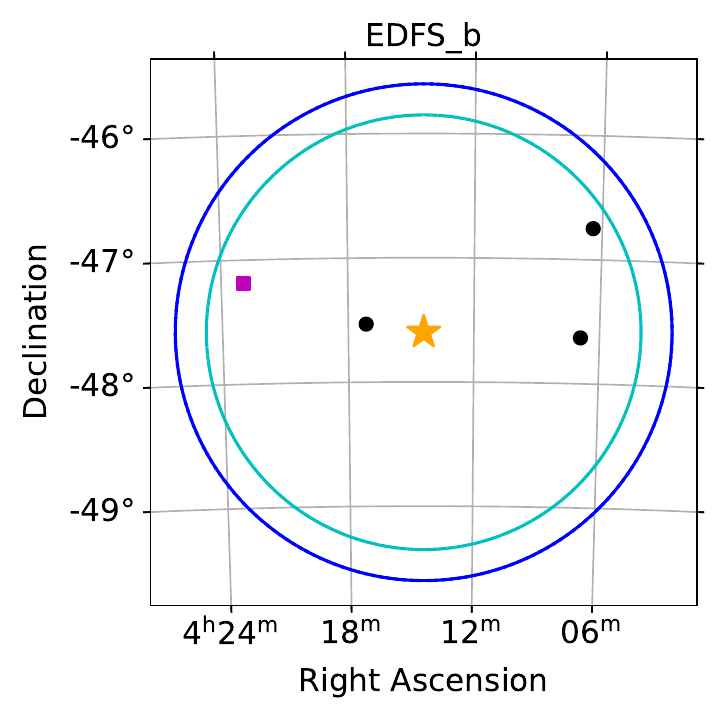}
  \plottwo{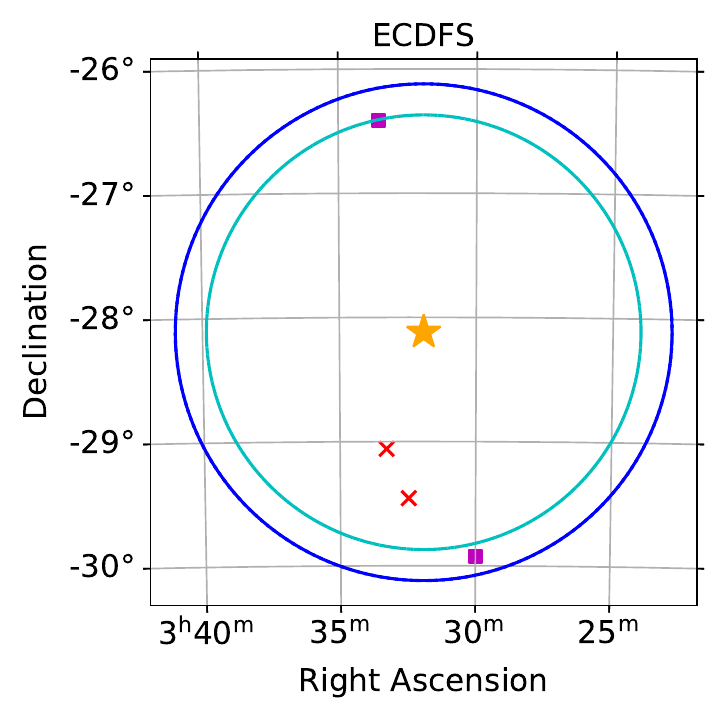}{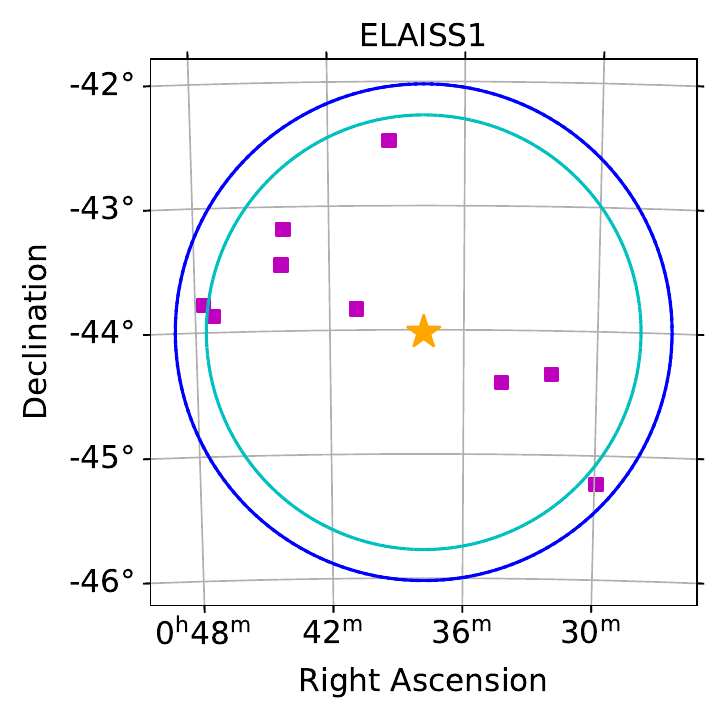}
  \caption{Aitoff projection of the six fields listed in Table \ref{tab1}. The star symbols mark the center of each field. The cyan and blue circles have a radius of 1.75~degree and 2.0~degree, representing the FOV of LSSTcam and the search radius, respectively. Red crosses, magenta squares and black points represent the RRL candidates without alerts, alerts with a small number of detections, and the alerts for the final candidates, respectively.}
  \label{fig_sky}
\end{figure*}

As of 08 May 2026, alerts have been produced in the Rubin DDFs plus the M49 field (i.e., the Virgo cluster, which was showcased during the Rubin First Look Event). We first queried selected RRL catalogs in these fields, using a search radius of 2~degrees centered on the equatorial coordinates listed in Table \ref{tab1}. We selected the same RRL catalog as in \citet{feng2024}, except that we used an updated DES RRL catalog from \citet{stringer2021}. In the case of the PS1 and DES RRL catalogs, we selected all candidates for RRL within the search areas, regardless of the probability or scores of the RRL classification given in these catalogs. The resulting numbers of RRL candidates in each field are listed in the fifth column of Table \ref{tab1}. We excluded RRL candidates with mean $g$-band magnitudes brighter than 16~mag, as they are likely to be saturated in the LSST camera \citep[LSSTcam,][]{lsstcam} images. The remaining number of candidates for the RRL are listed in the sixth column of Table \ref{tab1}. 

Using the {\tt ALeRCE} \citep[Automatic Learning for the Rapid Classification of Events,][]{forster2021} alert broker, we crossmatched the selected RRL candidates with the available alerts and extracted the available multiband light curves. For each alert, the number of detections across all bands ($N_{\mathrm{det}}$) is provided as {\tt nDiaSources}. In this pilot study, we excluded alerts with $N_{\mathrm{det}} < 50$, as their multiband lightcurves are too sparse to perform meaningful analysis. In columns 7, 8, and 9 of Table \ref{tab1}, we summarize the numbers of alerts with $1 < N_{\mathrm{det}} < 50$, the numbers of RRL candidates without alerts, and the number of alerts in the final sample (which will be analyzed in the next Section), respectively. Their locations in the sky are shown in Figure \ref{fig_sky}.

\subsection{Candidates without Alerts}

Since RRL are large amplitude pulsating stars, with $g$-band amplitudes as large as $\sim 1$~mag, these should be detected in the alerts via the difference imaging techniques. Therefore, the null detection of a given RRL candidate in alerts would hint at the non-variable nature of the candidate. Since the field-of-view (FOV) of LSSTcam has a radius of $\sim 1.75$~degrees, some of the RRL candidates located near the edge of a given field might not yet have alerts generated. Therefore, for the purpose of identifying candidates without alerts, we restricted those RRL candidates located within a radial separation of 1.75~degrees from the center of each field. In total, there are 9 RRL candidates without alerts in our sample which we present in Table \ref{tab_noalert}. In this Table, the mean $gri$-band magnitudes are adopted from the respective RRL catalogs.

Most of the candidates in Table \ref{tab_noalert} have relatively low scores (see footnote b in Table \ref{tab_noalert} for the definitions of the scores), suggesting that they are not RRL. However, they do not have early Rubin alerts generated, which could be due to the lack of template images for them or other observational constraints in this early data. Further verification or falsification of their nature has to await for future Rubin-LSST data.

\section{Alerts with Adequate Light Curves Data} \label{sec_analy}

\begin{deluxetable*}{lrlcccccccccc}
  %\movetableright=-1in
  \label{tab_cand}
  \tabletypesize{\scriptsize}
  \tablecaption{Alerts Associated with RR Lyrae Candidates.}
  \tablewidth{10pt}
  \tablehead{
    \colhead{Field} &
    \colhead{Alert OID} &
    \colhead{$N_{\mathrm{det}}$} &
    \colhead{$\Delta \mathrm{R}$ (deg.)} &
    \colhead{Catalog/ID} & 
    \colhead{Scores} &
    \colhead{$\langle g \rangle$} &
    \colhead{$\langle r \rangle$} &
    \colhead{$\langle i \rangle$} &
    \colhead{{\tt ALeRCE}\tablenotemark{a}} &
    \colhead{{\tt Lasair}} &
    \colhead{ST\tablenotemark{b}} &
    \colhead{RRL?} \\
    \colhead{(1)} &
    \colhead{(2)} &
    \colhead{(3)} &
    \colhead{(4)} &
    \colhead{(5)} &
    \colhead{(6)} &
    \colhead{(7)} &
    \colhead{(8)} &
    \colhead{(9)} &
    \colhead{(10)} &
    \colhead{(11)} &
    \colhead{(12)} &
    \colhead{(13)} 
  }
  \startdata
  \multicolumn{12}{c}{Case 1: Three ``VS'' in all three classifications} \\
EDFS\_b & 313681913612599326 & 135 & 0.468 & DES/1618003185 & 1.000/0.944/1.000 & 16.29 & 16.02 & 15.91 & VS(0.951) & VS & $\cdots$& Y \\
M49  & 170028526585511946 & 356 & 1.186 & PS1/118261861890373475 & 0.99/0.00 & 16.55 & 16.37 & 16.33 & VS(0.950) & VS & $\cdots$& Y \\
M49  & 170028526786838607 & 245 & 0.976 & PS1/116701878796340403 & 0.99/0.00 & 16.47 & 16.30 & 16.25 & VS(0.939) & VS & $\cdots$& Y \\
M49  & 170028538463780891 & 60  & 1.212 & PS1/118701864527905529 & 0.01/0.97 & 16.94 & 16.85 & 16.87 & VS(0.926) & VS & A0 & Y \\ 
M49  & 170028526583939107 & 299 & 1.341 & PS1/118431860912926265\tablenotemark{c} & 1.00/0.00 & 18.02 & 17.83 & 17.79 & VS(0.925) & VS & F0 & Y \\ 
COSMOS & 313976545432567829 & 672 & 0.951 & PS1/109651505272951351 & 0.92/0.01 & 16.47 & 16.29 & 16.25 & VS(0.902) & VS & $\cdots$& Y \\
M49  & 170028526719729676 & 512 & 0.911 & PS1/117861863572126768 & 1.00/0.00 & 17.19 & 17.00 & 16.97 & VS(0.897) & VS & $\cdots$& Y \\
COSMOS & 313967766913679434 & 345 & 0.945 & PS1/111601506486278284 & 0.99/0.00 & 16.01 & 15.85 & 15.83 & VS(0.874) & VS  & F0  & Y \\ 
M49  & 170028527032205382 & 300 & 0.436 & PS1/118101871246461596 & 0.99/0.00 & 17.15 & 16.96 & 16.88 & VS(0.786) & VS & $\cdots$ & Y \\
\multicolumn{12}{c}{Case 2: Two ``VS'' in all three classifications} \\
EDFS\_a & 313893022776427253 & 208 & 1.564 & DES/1436066478 & 1.000/0.722/1.000 & 16.86 & 16.58 & 16.59 & VS(0.955) & VS & $\cdots$& Y \\
COSMOS & 313888627082919999 & 553 & 0.864 & PS1/109641501706632370 & 0.99/0.00 & 18.81 & 18.64 & 18.61 & VS(0.947) & AGN & A0 & Y \\ 
COSMOS & 313888627074007083 & 359 & 1.356 & PS1/109541491369512425 & 0.74/0.02 & 16.91 & 16.70 & 16.65 & VS(0.946) & VS & $\cdots$& Y \\
COSMOS & 313879819412046118 & 276 & 1.716 & PS1/112021488127886394 & 0.99/0.00 & 17.39 & 17.18 & 17.13 & VS(0.944) & AGN & $\cdots$& Y \\
EDFS\_a & 313637935723315254 & 262 & 0.890 & DES/1440079396 & 1.000/1.000/0.986 & 16.27 & 16.03 & 16.12 & VS(0.937) & AGN & $\cdots$& Y \\
M49  & 170028526485897275 & 466 & 1.398 & NGVS/J122456.22+070149.2 & $\cdots$ & 23.31 & $\cdots$ & 22.45 & AGN(0.930) & VS & $\cdots$& N \\ 
M49  & 170028526449721369 & 386 & 1.813 & PS1/117811854281749298 & 0.99/0.00 & 17.97 & 17.75 & 17.68 & VS(0.928) & AGN & $\cdots$& Y \\
M49  & 170028526533607595 & 377 & 1.680 & PS1/115631876364037111 & 0.61/0.04 & 17.63 & 17.39 & 17.32 & VS(0.903) & VS & $\cdots$& Y \\
EDFS\_b & 313677516285411354 & 407 & 1.263 & DES/1457011513 & 1.000/0.944/0.999 & 17.73 & 17.48 & 17.38 & VS(0.894) & AGN & $\cdots$& Y \\
EDFS\_a & 170028485771788495 & 125 & 1.640 & DES/1436029179 & 1.000/0.722/0.999 & 17.07 & 16.74 & 16.66 & VS(0.795) & VS & $\cdots$& Y \\
M49  & 170028527324233872 & 133 & 0.503 & PS1/117661877543738692\tablenotemark{c} & 0.99/0.00 & 18.68 & 18.52 & 18.48 & VS(0.786) & AGN & $\cdots$ & Y \\
COSMOS & 313985345023639575 & 774 & 0.514 & PS1/111021496859276457 & 0.92/0.02 & 17.90 & 17.68 & 17.62 & VS(0.689) & AGN & F0 & Y \\ 
M49  & 170028526470168653 & 419 & 1.046 & PS1/117291862253949298\tablenotemark{c} & 1.00/0.00 & 18.78 & 18.59 & 18.54 & VS(0.688) & AGN & $\cdots$& Y \\
COSMOS & 313853517411385444 &1004 & 0.970 & PS1/111481508121971071 & 0.44/0.01 & 18.50 & 18.34 & 18.33 & VS(0.636) & VS  & F5 & N \\ 
M49  & 170028526620639344 & 495 & 1.924 & PS1/116241856788715321 & 0.92/0.01 & 18.56 & 18.35 & 18.31 & VS(0.625) & AGN &  $\cdots$& Y \\
COSMOS & 313853518024802332 & 124 & 1.232 & PS1/109191501512579201 & 0.97/0.00 & 19.40 & 19.19 & 19.13 & AGN(0.593) & VS & $\cdots$ & N \\ 
\multicolumn{12}{c}{Case 3: One ``VS'' in all three classifications} \\
EDFS\_a & 313765480573698107 & 362 & 0.530 & DES/1425463757\tablenotemark{d} & 1.000/0.722/1.000 & 17.93 & 17.78 & 17.76 & VS(0.938) & AGN & $\cdots$& Y \\
EDFS\_a & 170028486401458264 & 50 & 1.687 & DES/1420085264 & 1.000/0.722/1.000 & 17.56 & 17.32 & 17.43 & VS(0.904) & AGN & $\cdots$& Y \\
M49  & 170028526474363023 & 504 & 0.754 & PS1/117641864896623415\tablenotemark{c} & 0.99/0.00 & 19.10 & 18.88 & 18.82 & bogus(0.811) & AGN &  $\cdots$& Y \\
M49  & 170028527203647571 & 462 & 1.488 & PS1/115851875605203450\tablenotemark{c} & 1.00/0.00 & 18.73 & 18.53 & 18.48 & AGN(0.791) & AGN & A0 & Y \\ 
COSMOS & 170028511812124695 & 115 & 0.955 & PS1/110861491668282127 & 0.40/0.01 & 19.09 & 18.79 & 18.69 & bogus(0.756) & VS & $\cdots$ & N \\ 
EDFS\_b & 313681913543393322 & 338 & 1.599 & DES/1454783451 & 1.000/0.500/0.999 & 19.33 & 19.11 & 19.10 & VS(0.729) & AGN & $\cdots$& Y \\ 
M49  & 170028526486945938 & 561 & 1.756 & PS1/115851862640860700\tablenotemark{c} & 1.00/0.00 & 19.00 & 18.82 & 18.81 & AGN(0.712) & AGN & A1 & Y \\ 
M49  & 170028526607007832 & 349 & 0.542 & PS1/118041868502554612\tablenotemark{c} & 0.97/0.01 & 18.72 & 18.55 & 18.52 & AGN(0.697) & AGN & F5 & Y \\ 
M49  & 170028526897987638 & 299 & 0.168 & NGVS/J122940.52+080057.2 & $\cdots$ & 19.21 & $\cdots$ & 19.07 & AGN(0.667) & AGN & $\cdots$& Y \\
COSMOS & 313936986529333309 & 759 & 1.352 & PS1/109051501245213565 & 0.03/0.83 & 16.93 & 16.93 & 16.99 & AGN(0.633) & VS & A0 & ? \\ 
M49  & 170028527591620680 & 365 & 0.554 & NGVS/J122743.96+073237.4 & $\cdots$ & 18.92 & $\cdots$ & 18.75 & AGN(0.573) & AGN & $\cdots$& N \\ 
  \multicolumn{12}{c}{Case 4: No ``VS'' in all three classifications} \\
COSMOS & 170028511005769863 & 159 & 1.622 & PS1/110851484938128042 & 0.12/0.01 & 19.80 & 19.58 & 19.54 & AGN(0.872) & AGN & QSO & N \\ 
M49  & 170028532667252850 & 469 & 0.886 & PS1/117101864600112191 & 0.36/0.02 & 18.01 & 17.87 & 17.85 & AGN(0.837) & AGN & $\cdots$& ? \\
COSMOS & 170028500595507341 & 923 & 1.486 & PS1/108941497491486712 & 0.02/0.01 & 18.60 & 18.40 & 18.34 & AGN(0.822) & AGN & QSO & N \\ 
M49  & 170028526517354584 & 467 & 1.175 & PS1/116241875890218931\tablenotemark{c} & 0.04/0.01 & 20.37 & 20.14 & 20.10 & bogus(0.797) & AGN & $\cdots$& N \\
  \enddata
  \tablenotetext{a}{Values in the parenthesis are the associated probability for the classification returned from {\tt ALeRCE}. The adopted classifier version is {\tt stamp\_classifier\_rubin\_beta\_20260421}.}
  \tablenotetext{b}{Spectrum type (ST) retrieved from using the {\tt AstroInspect} \citep{cardoso2026} web-tool.}
  \tablenotetext{c}{These PS1 RRL candidates were also in the NGVS RRL catalogs.}
  \tablenotetext{d}{This RRL was also observed in DP1, with the name of  SSS J035520.3-484728.}
  \end{deluxetable*}

There are 40 candidates with $N_{\mathrm{det}} \geq 50$ listed in Table \ref{tab_cand}. {\tt ALeRCE} has included a beta-version of the stamp-based classifier \citep[which is similar to the one described in][but updated for the Rubin Observatory alerts]{cd2021} to classify the alerts into five broad categories: AGN, variable stars (VS), supernovae (SN), asteroids and bogus. The stamp-based classifier has also provided classification probabilities. In addition to {\tt ALeRCE}, we have also included the classification of each alert from the {\tt Lasair} \citep{williams2024} alert broker which classifies the alerts using an independent classifier. {\tt Lasair} performs a contextual classification of alerts by cross-matching them with a number of catalogs using the {\tt Sherlock} \citep{Young_sherlock_2023} framework. {\tt Sherlock} classifies alerts into seven broad categories, including VS and AGN, and five other categories associated with transients. The classifications of the 40 alerts from {\tt ALeRCE} and {\tt Lasair} are summarized in columns 10 and 11, respectively, of Table \ref{tab_cand}.

\subsection{Assessment Based on Classifications}

In total, there are three classifications available for the RRL candidates listed in Table \ref{tab_cand}, including the scores of the original RRL catalogs and the classifications provided by the {\tt ALeRCE} and {\tt Lasair} alert brokers. In the PS1 RRL catalog, two scores for RRL candidates were given as $S3ab$ and $S3c$, representing the ab-type (or fundamental mode) and c-type (for first-overtone) RRL, respectively. Therefore, we adopted the score as $S_{PS1}=\mathrm{max}(S3ab,\ S3c)$ for a given RRL candidate. For the candidate in the DES RRL catalog, we follow the approach of \citet{feng2026} by assigning a given candidate a score $S_{DES} >0.9$ if all three RF (random-forest) scores are greater than 0.9. The RRL candidates in the NGVS RRL catalogs already have a score of 0.93 or greater \citep{feng2024}. Then, a RRL candidate that has a score larger than 0.9 from any of these RRL catalogs would be treated as ``VS'' (to be consistent with alert-based classifications). Together, there are four cases based on these three classifications, as described in further detail in the following. 

{\it Case 1: All three classifications suggest that an RRL candidate is ``VS''.} There are nine RRL candidates in this case and are confidently RRL stars.

{\it Case 2: Two of the classifications classified an RRL candidate as ``VS''.} Based on these classifications, the candidates are most likely RR Lyrae. Nevertheless, we have found that a small number of candidates are actually not RRL based on their multiband light curve shapes (see next subsection).

{\it Case 3: Only one of the classifications resulted in an RRL candidate as ``VS''.} The candidates in this case are most likely not RRL. However, we found that many of the candidates are indeed RRL based on light curve morphology (see next subsection).

{\it Case 4: None of the classifications suggest an RRL candidate is ``VS''.} The four candidates have quite a low score from the PS1 RRL catalogs and were classified as AGN or bogus from the other two classifiers. They are likely not RRL variables.

\begin{figure}
  \epsscale{1.1}
  \plotone{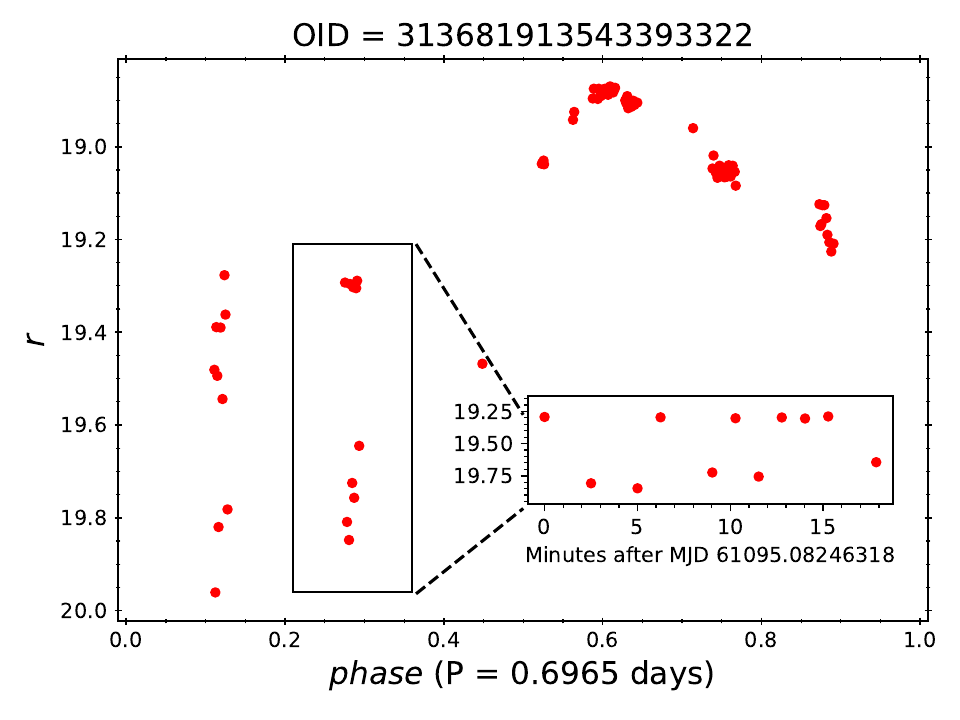}
  \caption{The $r$-band light curve for an alert associated with an RRL from the DES RRL catalog. The light curve has been folded using its pulsation period of 0.6965~days. The inset figure shows the intranight variability for data points highlighted in the box.}
  \label{fig_demo}
\end{figure}

\subsection{Constraint from Alert Light Curves}

Using {\tt ALeRCE}, we downloaded the multiband alert light curves for the candidates listed in Table \ref{tab_cand}, where the data points on the light curves were converted from $scienceFlux$ to AB magnitudes using $m=31.4-2.5\log_{10} scienceFlux$, and the error is calculated as $\sigma_m = (2.5/\log 10) \times (scienceFlux\_error/scienceFlux)$. A cautionary note on these light curves from early Rubin alerts is that some of them exhibit intranight variations\footnote{This was pointed out by one of the authors (SGK), and was highlighted within the LSST Stars, Milky Way and Local Volume Science Collaboration (2026 - private communication) and in the Rubin community forum (see \url{https://community.lsst.org/t/photometric-quality-of-alert-photometry/11720}).} within short time scales (typically less than 20 minutes). Such variations can be as large as $\sim0.5$~mag and are likely not related to the intrinsic variability of the source. A randomly selected example is shown in Figure \ref{fig_demo} for an alert with OID = 313681913543393322\footnote{The OID is the {\tt diaObjectId} associated with the alert.}, and we emphasize that in this figure, as well as in all subsequent figures, the error bars at the data-points are smaller than the size of the symbols.   

\begin{figure}
  \epsscale{1.1}
  \plotone{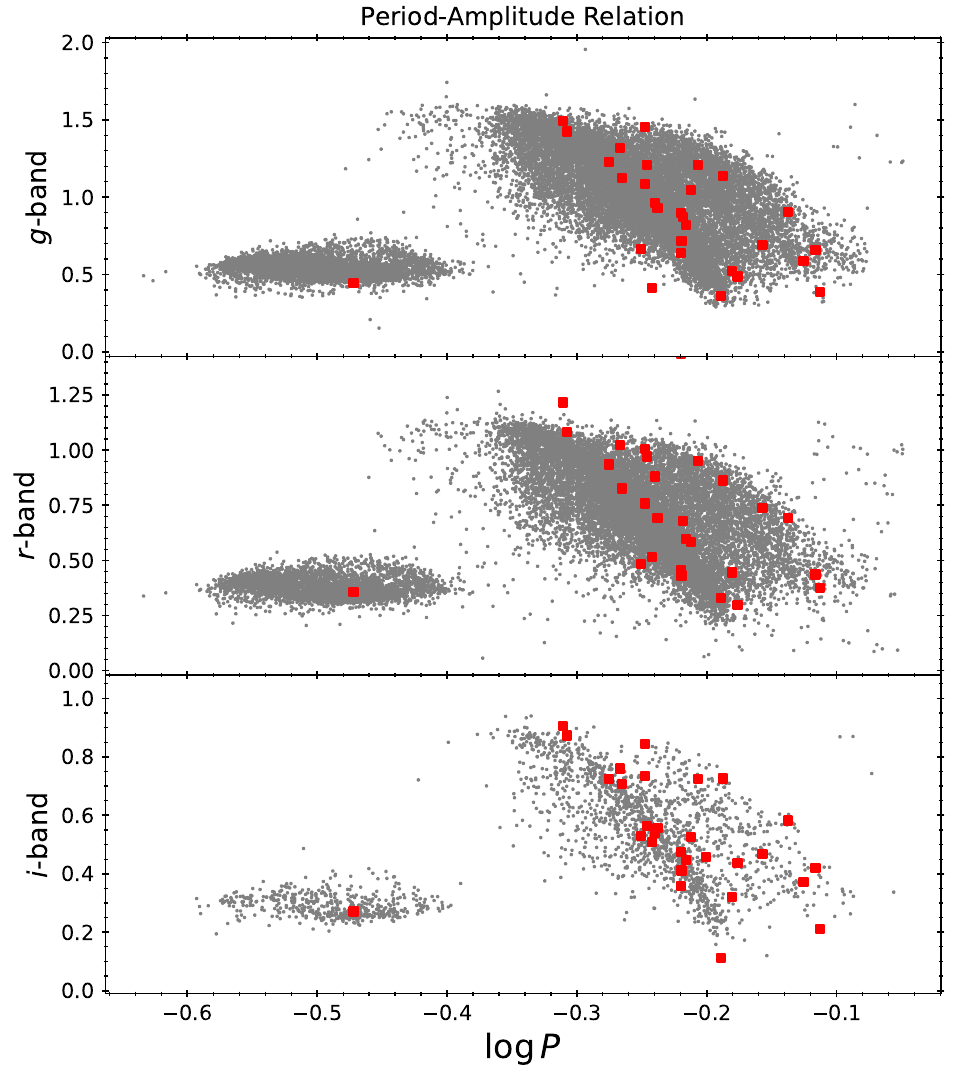}
  \caption{The $gri$-band amplitudes, shown as red squares, for the confirmed RRL in Case 1, 2 and 3 as listed in Table \ref{tab_cand} on the period-amplitude relation. These amplitudes were determined from fitting the template light curves \citep{braga2024} to the alert's light curves. However, due to the presence of intranight variability in the light curves, these amplitudes should be treated as {\it preliminary}. For comparisons, we have also included the amplitudes, shown as gray points, for $\sim 44,700$ known RRL adopted from \citet{braga2024}.} 
  \label{fig_pa}
\end{figure}

\begin{figure*}
  \epsscale{1.1}
  \plottwo{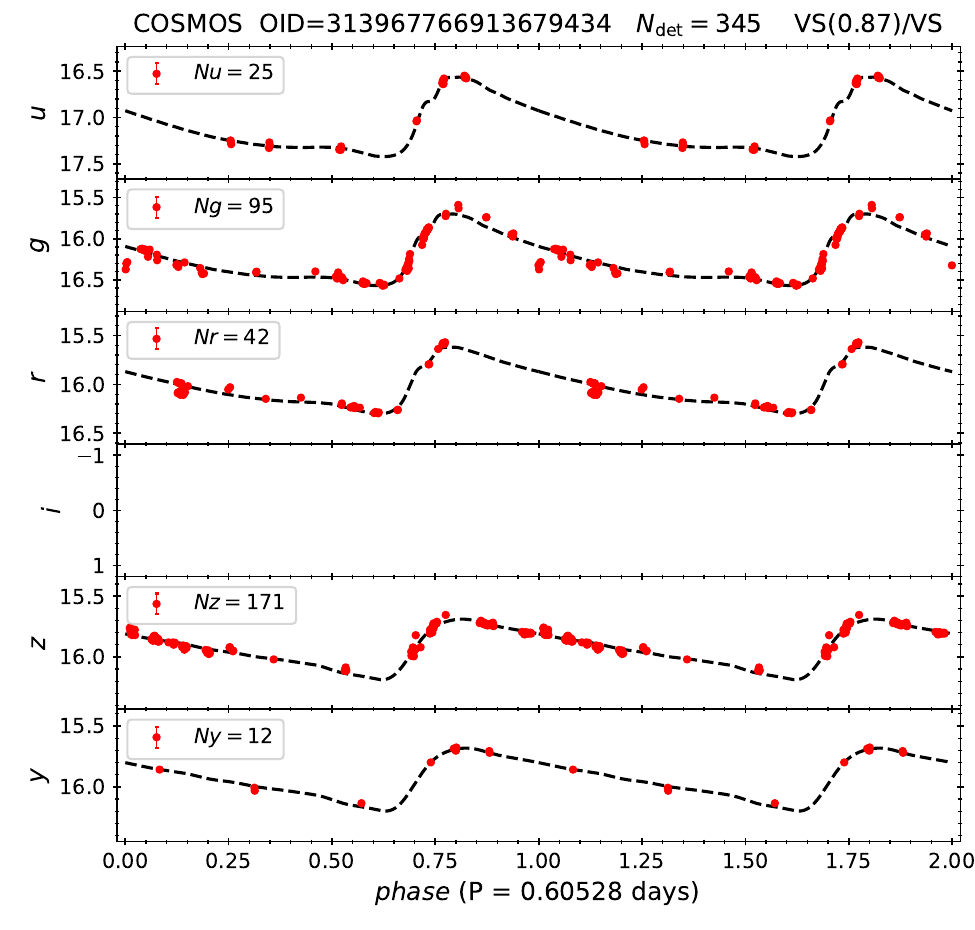}{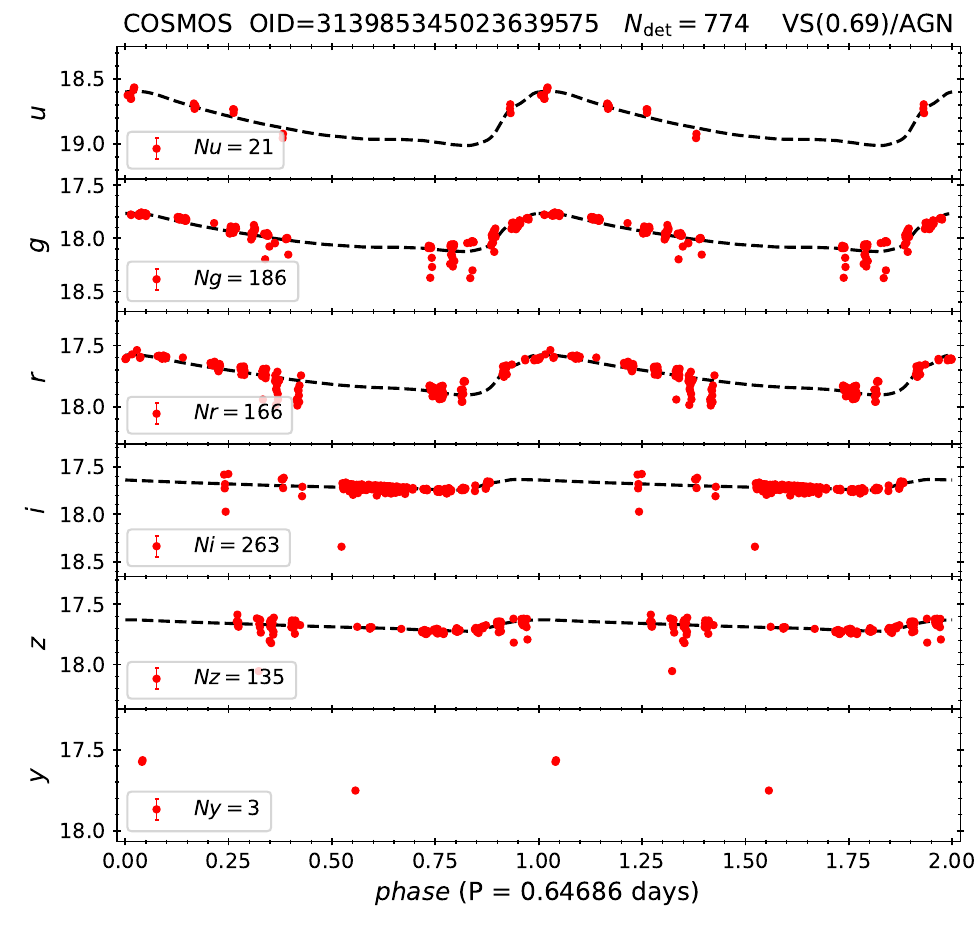}
  \caption{Multiband light curves for two confirmed RRL in the COSMOS field. The dashed curves are the best-fit template light curves \citep{braga2024}. Note that the COSMOS field is the only field that is observed in all six $ugrizy$ filters in the early Rubin alerts (even though not all RRL have six-band light curves). The \citet{braga2024} template light curves did not include the $u$- and $y$-band, therefore we fit the light curves data using the $g$- and $z$-band template light curves, respectively.}
  \label{fig_cosmos}
\end{figure*}

\begin{figure*}
  \epsscale{1.1}
  \plottwo{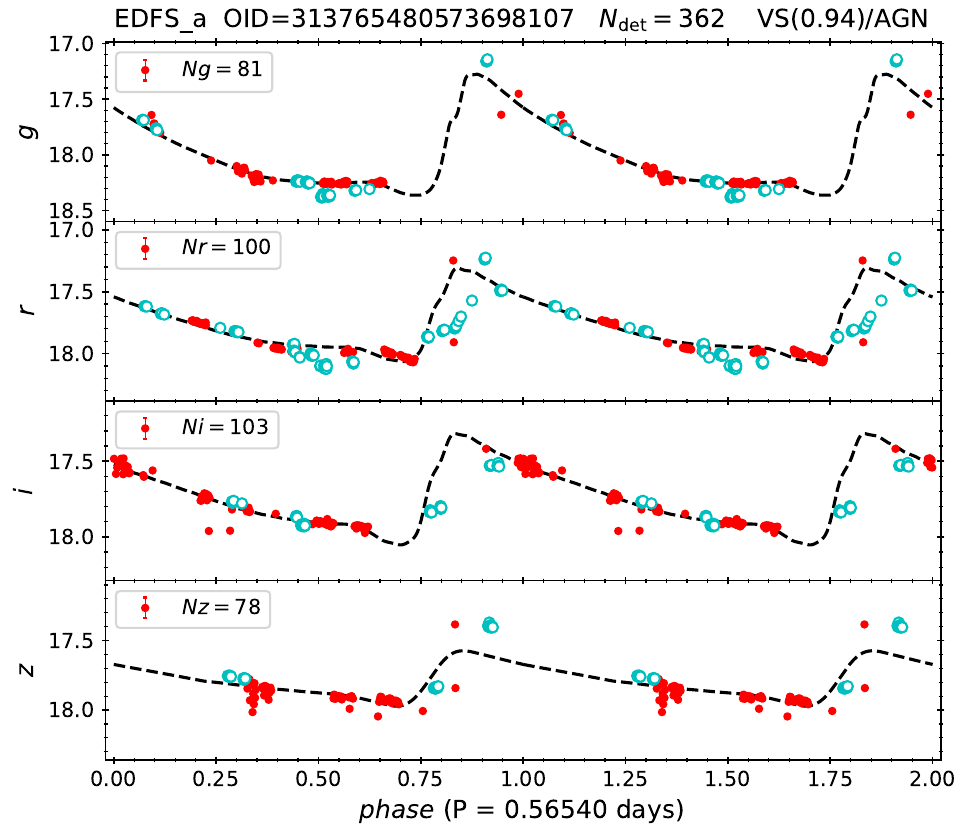}{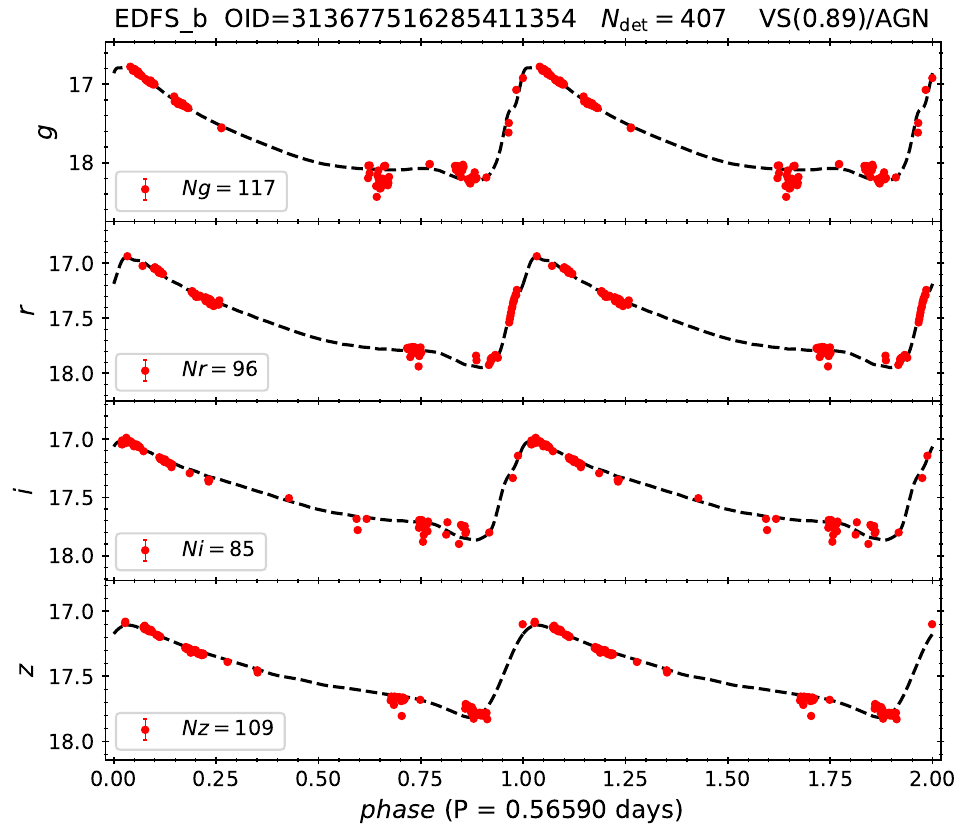}
  \caption{Same as Figure \ref{fig_cosmos}, but for two confirmed RRL in the EDFS\_a/b field. These fields have been observed in the $griz$ filters. The RRL candidate on the left panels has also been observed in the Rubin DP1, and the corresponding light curves are shown as open cyan circles. Note that this RRL is suspected to be a Blazhko RRL \citep{ngeow2026}.}
  \label{fig_edfs}
\end{figure*}

\begin{figure*}
  \epsscale{1.1}
  \plottwo{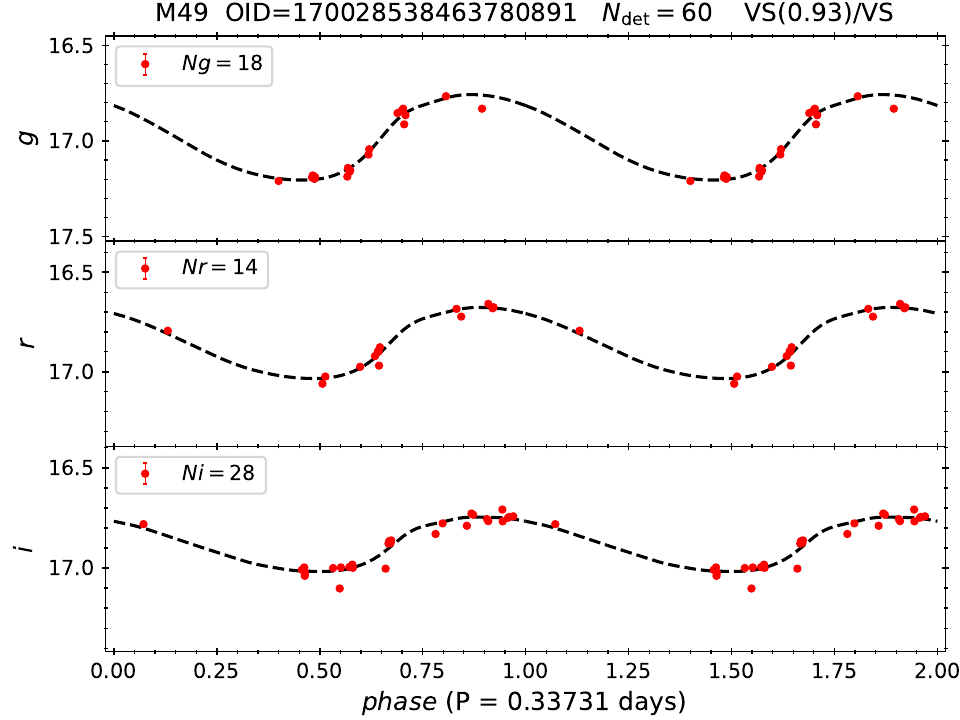}{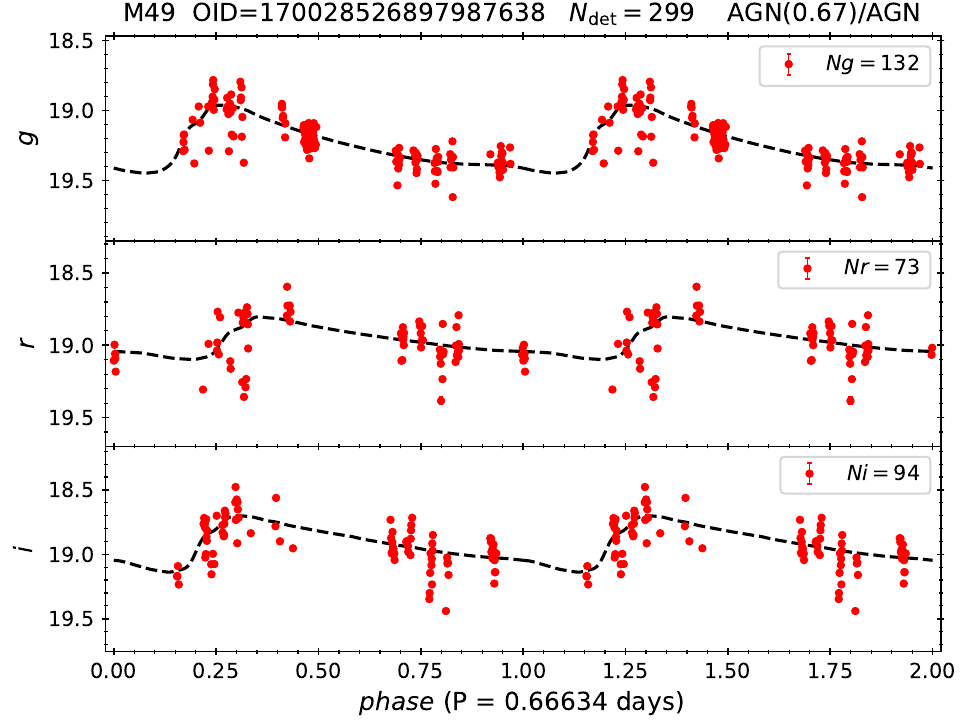}
  \caption{Same as Figure \ref{fig_cosmos}, but for two confirmed RRL in the M49 field, which was only observed in the $gri$-band. Note that the left panels show the light curve for a c-type RRL rather than the ab-type RRL (as in the right panels).}
  \label{fig_m49}
\end{figure*}

For each candidate, we folded their multiband light curves using their pulsation periods ($P$) adopted from their respective crossmatched PS1, DES, or NGVS RRL catalogs (the adopted catalogs are listed in column 5 of Table \ref{tab_cand})\footnote{The only overlapping catalogs for our 40 candidates are the PS1 and NGVS catalogs in the M49 field. Periods from both catalogs are in general agreement with each others \citep{feng2024}.}, and visually inspected these light curves to examine whether they resembled RRL-like light curves or not. Based on this visual inspection, we marked ``Y'' in the last column of Table \ref{tab_cand}, indicating that they are truly RRL, else we classified them as non-RRL and marked ``N'' in Table \ref{tab_cand}. There are two candidates that show ambiguous results. They are marked as ``?'' in Table \ref{tab_cand} and are discussed further in the next subsection. To aid the final classification, we included the spectral types for these candidates, where available, retrieved from the {\tt AstroInspect} \citep{cardoso2026} web-tool. As expected, all candidates in case 1 are genuine RRL. 

\begin{figure*}
  \epsscale{1.1}
  \plottwo{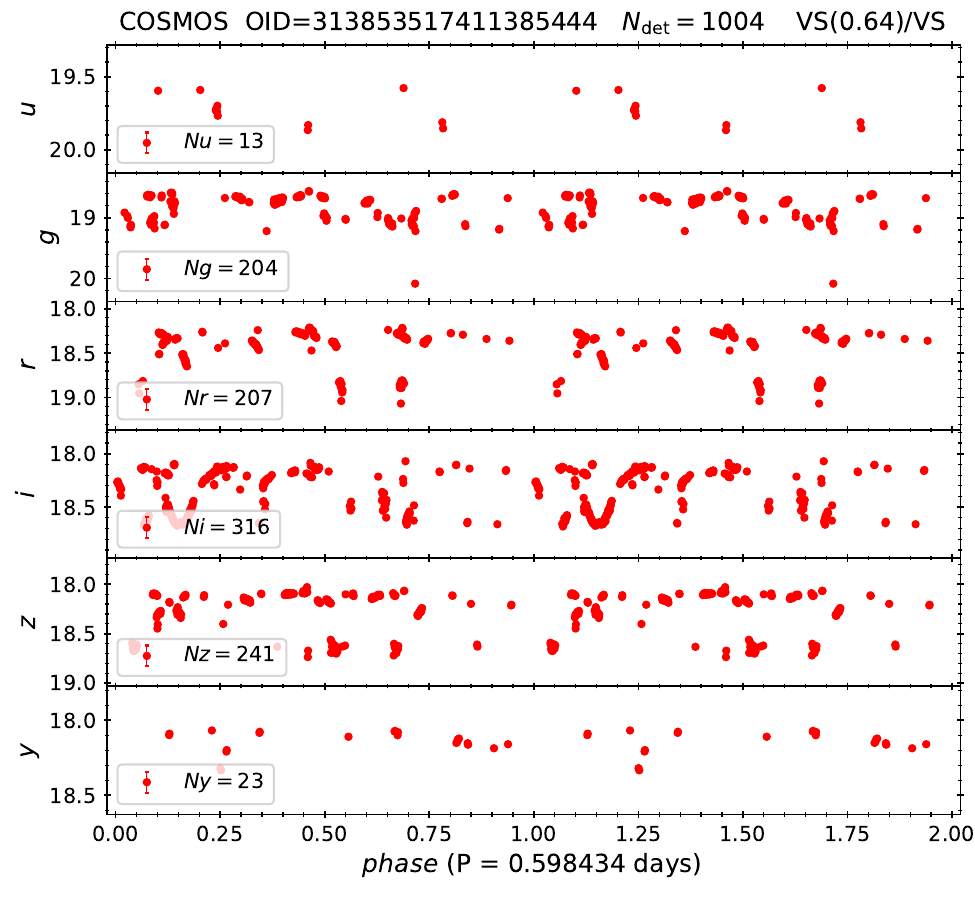}{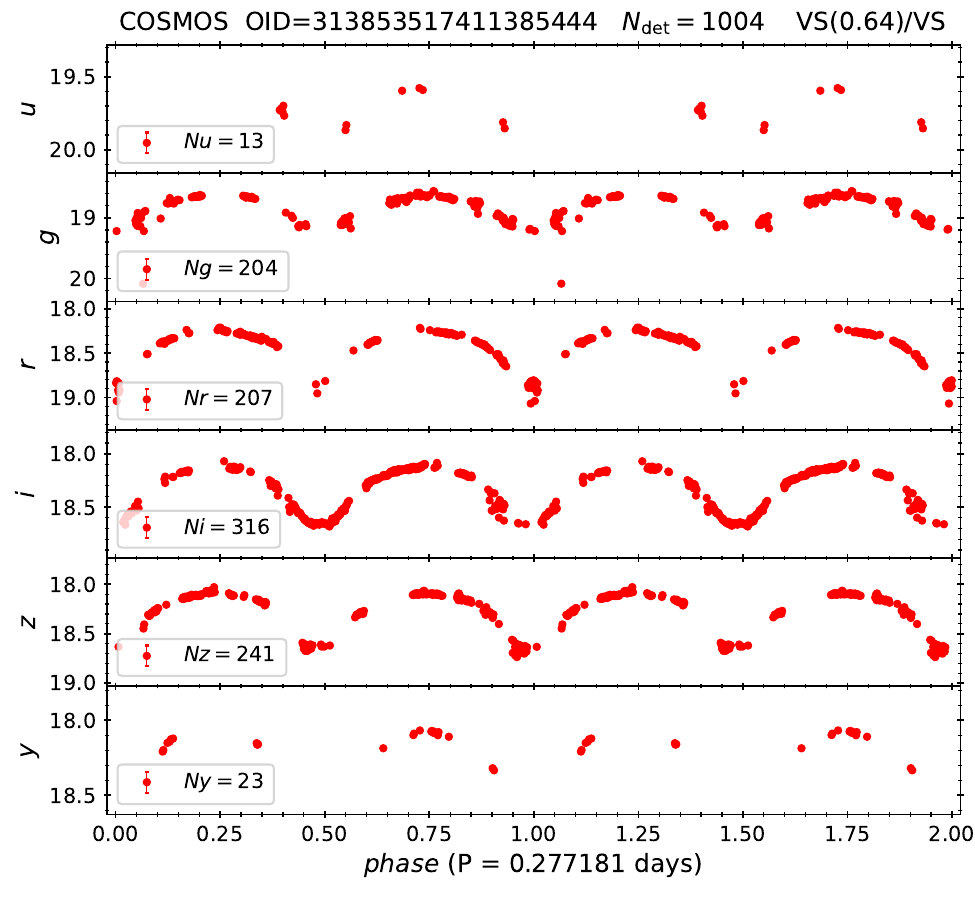}
  \caption{Multiband light curves for the alert (313853517411385444) associated with a known contact binary. Left and right panels are the light curves folded with periods given in the PS1 RRL catalog and in \citet{drake2014}, respectively.}
  \label{fig_pcosmos}
\end{figure*}

\begin{figure*}
  \epsscale{1.1}
  \plottwo{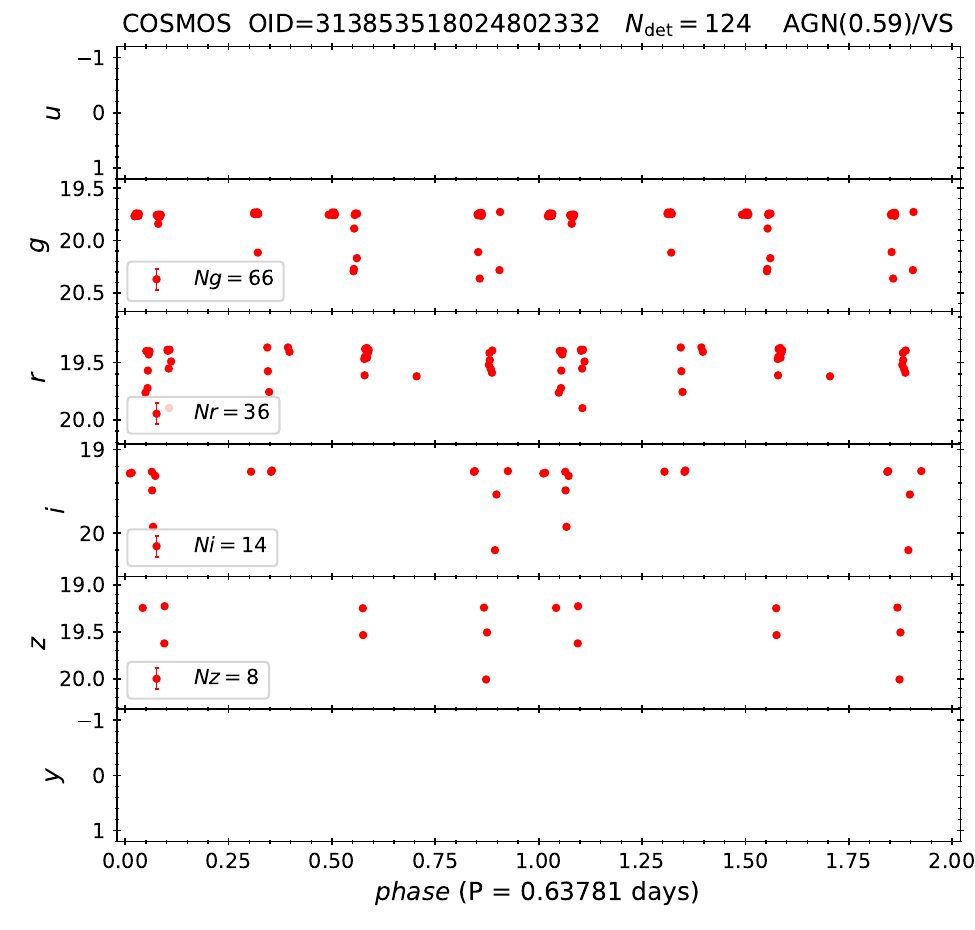}{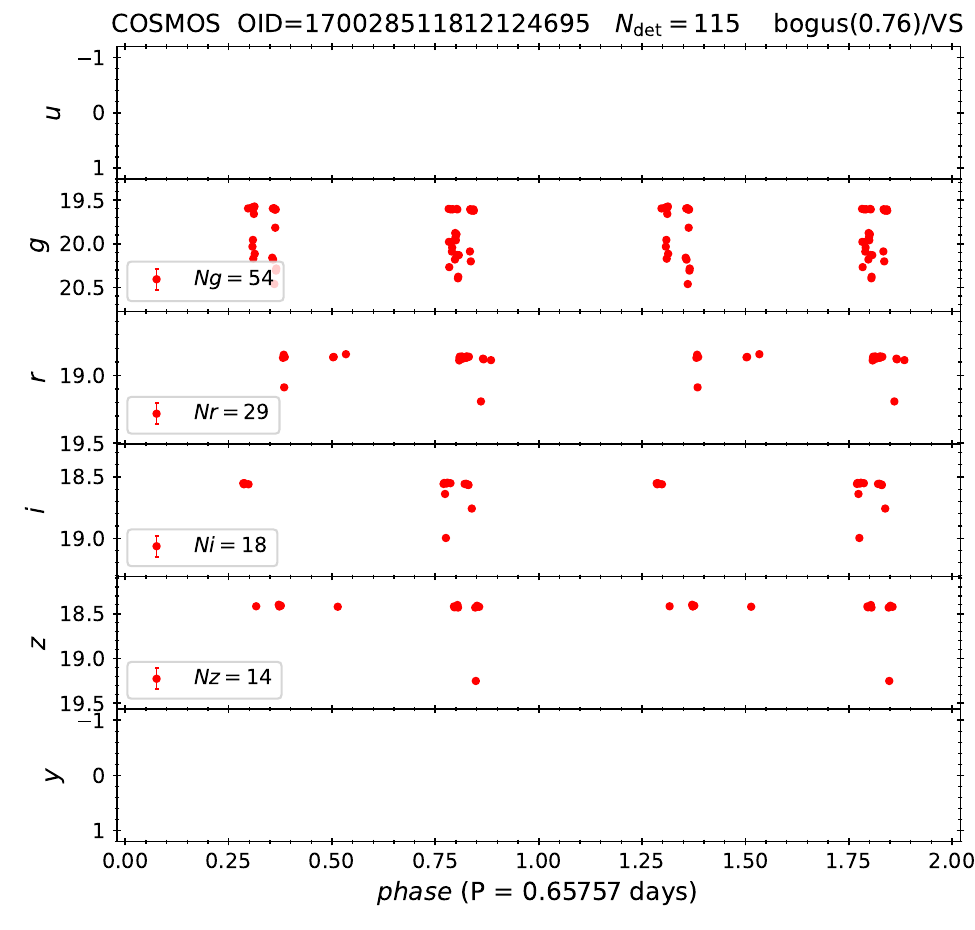}
  \plottwo{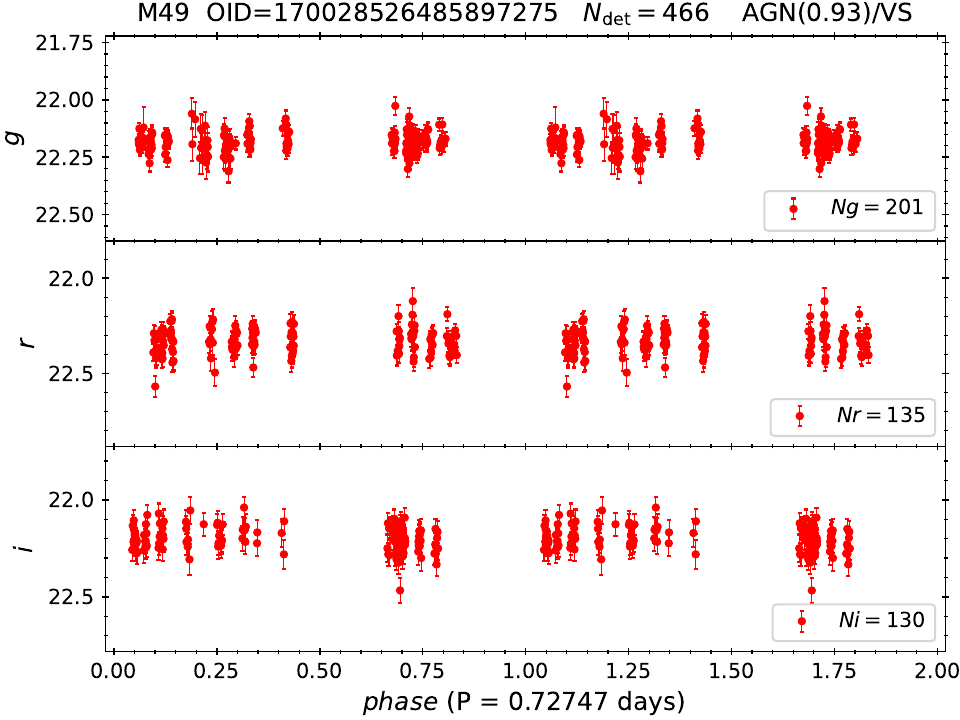}{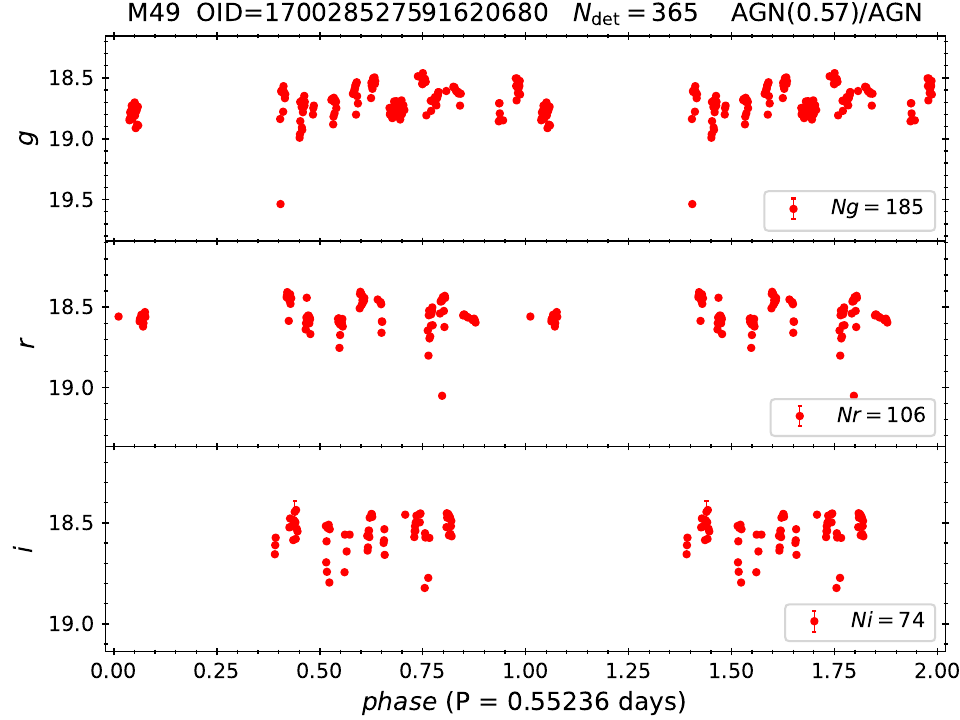}
  \caption{Multiband light curves for candidates in Cases 2 and 3, located in the COSMOS and the M49 field, that are not RR Lyrae. Note that it is unclear whether the large scatter seen in a given phase is intrinsic or due to the intranight variations.}
  \label{fig_no}
\end{figure*}

Using multiband alert light curves, the majority of candidates in Cases 2 and 3 were confirmed to be RRL. Their amplitudes matched the distribution of the known RRL in the period-amplitude diagram (see Figure \ref{fig_pa}, which also includes the RRL in Case 1). In Figures \ref{fig_cosmos}, \ref{fig_edfs} and \ref{fig_m49}, we present the multiband light curves for two randomly selected RRL (drawn from Cases 1 through 3) in the COSMOS, EDFS\_a / b and M49 fields, respectively. Note that one of the RRL in the EDFS field was also included in Rubin Data Preview 1 \citep[DP1,][]{RTN-095} and investigated by \citet{ngeow2026}. To guide the eyes, we have also fitted these light curves using the template light curves available from \citet{braga2024}. As can be seen from these figures, there is a mixture of light curve qualities. Some of them have very good light curve quality, while others exhibit the intranight variation mentioned earlier, with some outliers present in the light curves. Because of these, we do not attempt to derive their mean magnitudes in this study.\footnote{The photometric properties will be derived in future work with better quality light curves, which will be improved with the data taken during the official Rubin-LSST observations.}

\begin{figure*}
  \epsscale{1.1}
  \centering
  \begin{tabular}{ccc}
    \includegraphics[width=5.75cm]{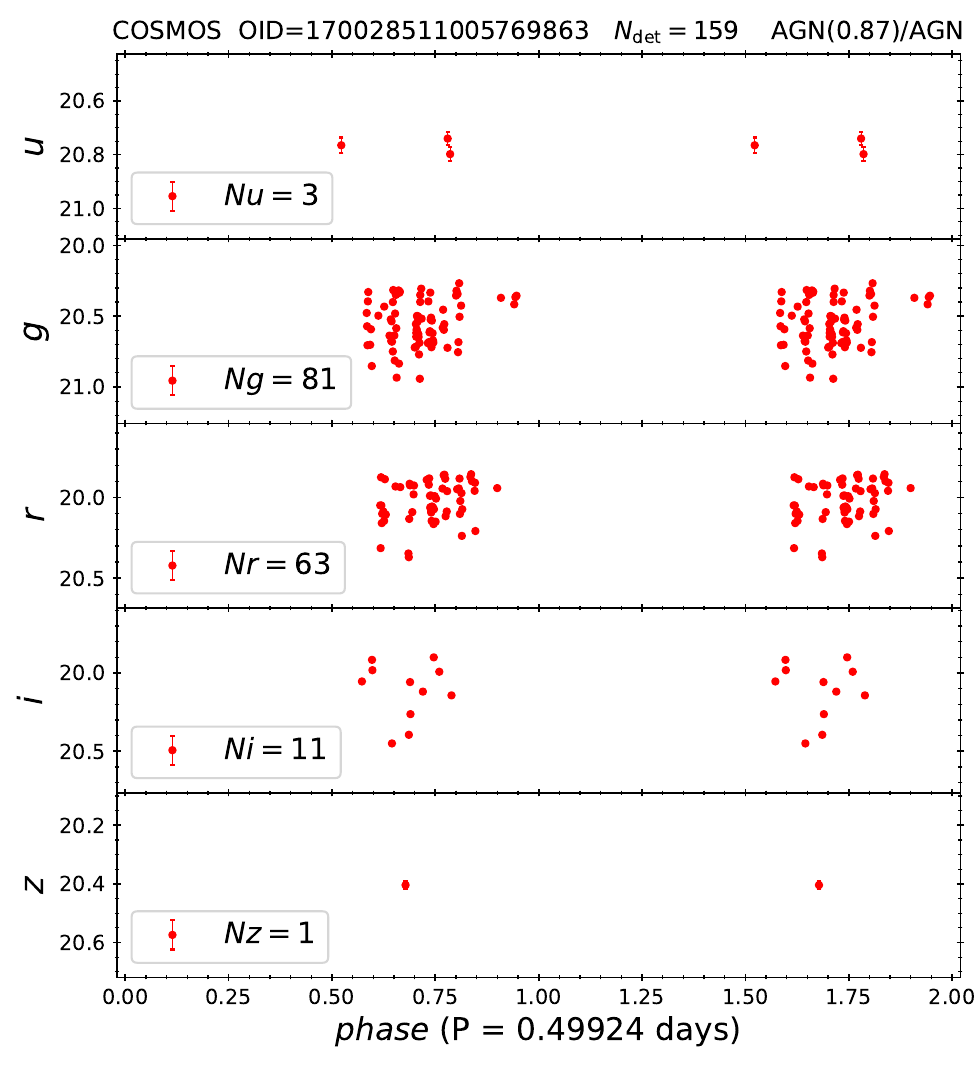} & \includegraphics[width=5.75cm]{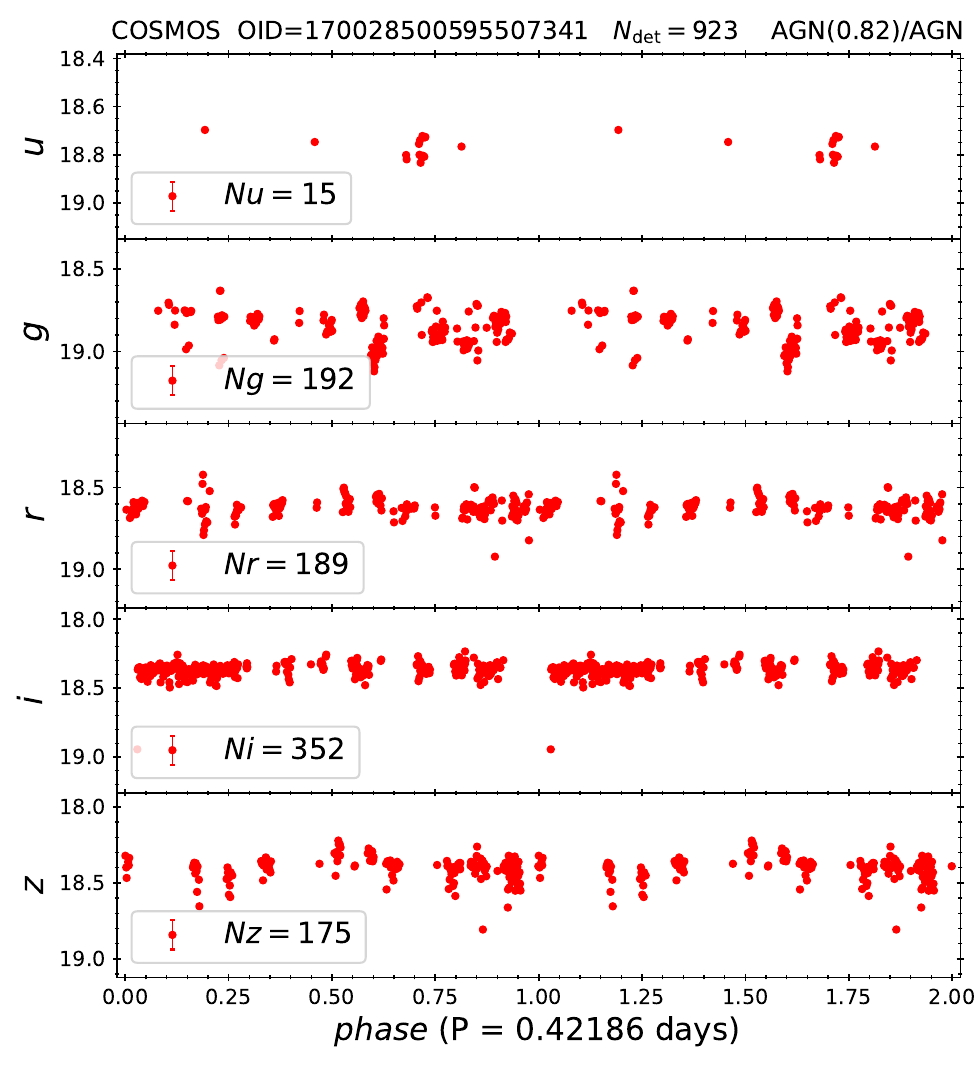} & \includegraphics[width=5.75cm]{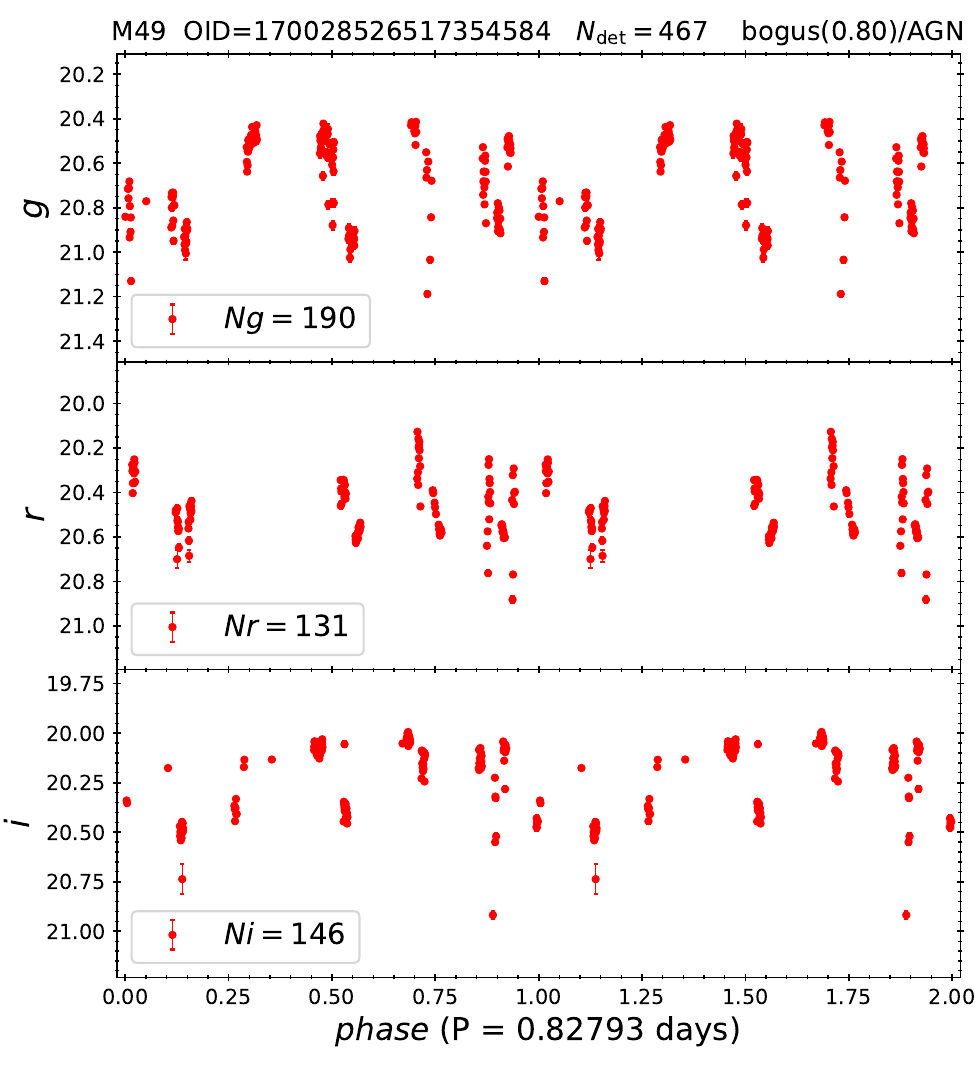} \\
  \end{tabular}
  \caption{The multiband light curves for the three candidates in Case 4, which do not exhibit light curve shapes expected for RRL. Hence, they are rejected as RRL candidates.}
  \label{fig_case4}
\end{figure*}

There are five candidates in Cases 2 and 3 that are rejected as RRL based on the multiband alert light curves. One of them is a known contact binary \citep{drake2014}. Using the correct period from \citet{drake2014}, the light curves resembled the contact binaries light curves, as demonstrated in the right panels of Figure \ref{fig_pcosmos}. For the other four rejected candidates, their light curves are shown in Figure \ref{fig_no}. Finally, three of the four candidates in Case 4 can be rejected as RRL based on their multiband light curves, as shown in Figure \ref{fig_case4}. Two of them are confirmed to be AGN on the basis of the available spectra. For alert 170028526517354584, even though \citet[][using time-series NGVS data]{feng2024} classified it as a c-type RRL, it has a very low score of $S_{PS1}=0.04$ from the PS1 RRL catalog. Based on random-forest models, \citet{arevalo2026} predicted that this candidate is a short-period eclipsing binary. 

\subsection{Ambiguous Candidates}

\begin{figure*}
  \epsscale{1.1}
  \plottwo{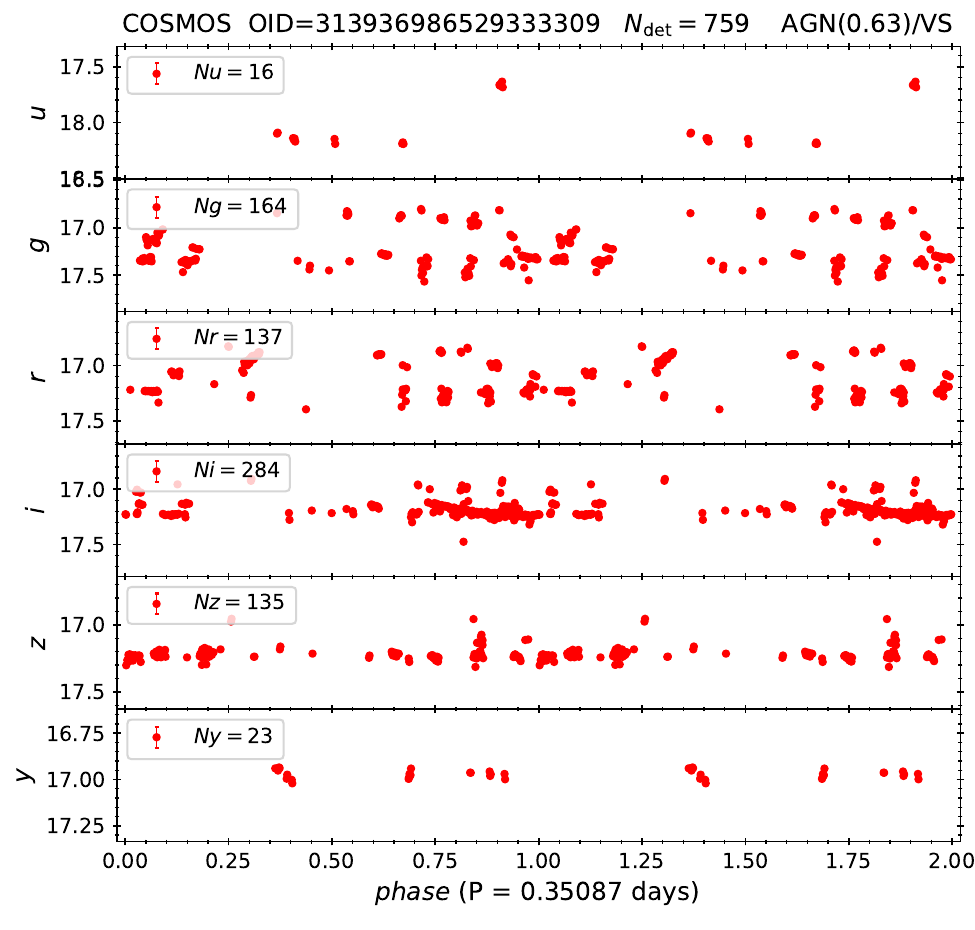}{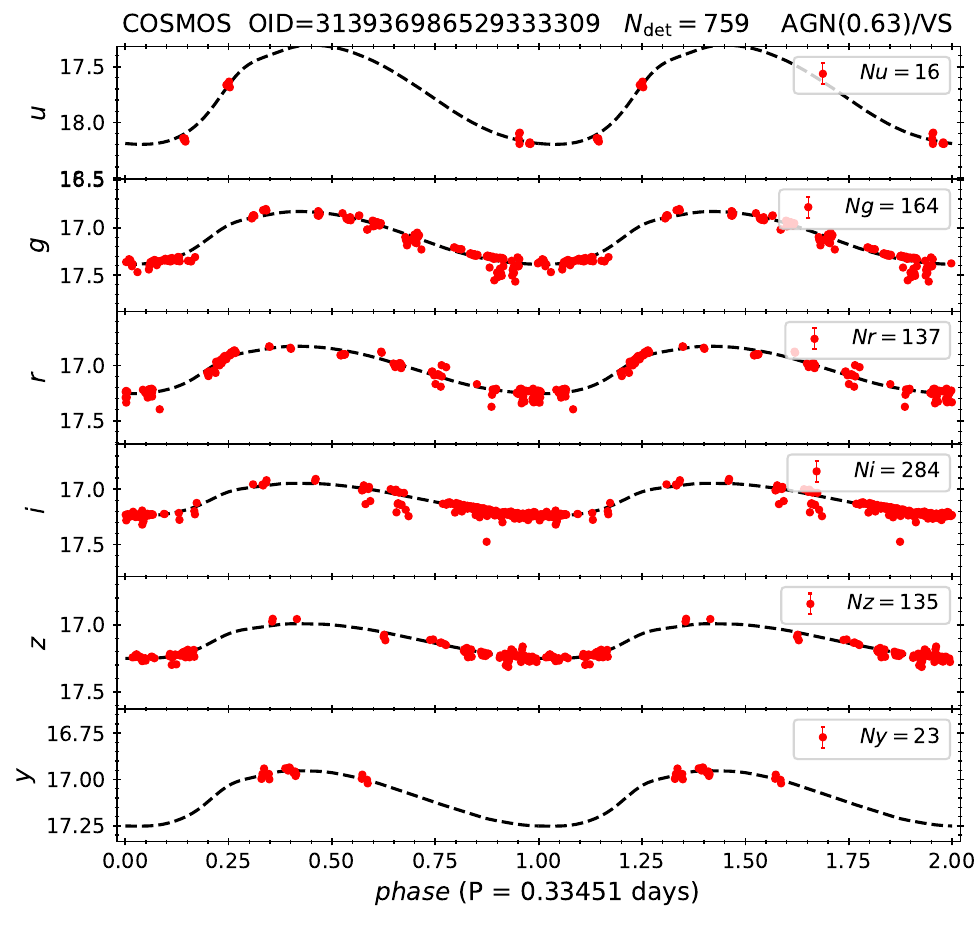}
  \caption{Folded multiband light curves for alert 313936986529333309. The left panels show the resulting light curves when folded with the default period adopted from the PS1 RRL catalog. Light curves in the right panels were folded with a shorter period adopted from \citet{chen2020} or \citet{clementini2023}. The dashed curves are the best-fit template light curves for a c-type RRL. Note that the \citet{braga2024} template light curves are only available in the $gri$-band for the c-type RRL. Therefore, we adopted the $g$-band and $i$-band template light curves for fitting the $u$-band and the $zy$-band light curve data, respectively.}
  \label{fig_3309}
\end{figure*}

The candidate associated with the alert 313936986529333309 is an interesting case. It has a marginal score of $S_{PS1}=0.83$ as a c-type RRL, and is classified as ``VS'' by one of the alert brokers. Furthermore, its spectral type is consistent as a RRL. In the literature, it was classified either as a c-type RRL \citep[][and in PS1 RRL catalog; with $P=0.33452$ and $0.35087$~days, respectively]{chen2020} or as an ab-type RRL \citep[][with $P=0.50336$~days]{drake2014}. Interestingly, the Gaia Data Release 3 (DR3) RRL catalog \citep{clementini2023} classified this candidate as an ab-type RRL but with a period of $0.33451$~days. As demonstrated in Figure \ref{fig_3309}, the period we adopted from the PS1 RRL catalog did not fold the alert multiband light curves well, whereas light curves folded with a shorter period fit well with the c-type RRL template light curves. The classification of this candidate as a c-type RRL and a shorter period of $P\sim0.335$~days was further supported by the work of \citet{chan2022} and \citet{ss2023}, respectively.

\begin{figure}
  \epsscale{1.1}
  \plotone{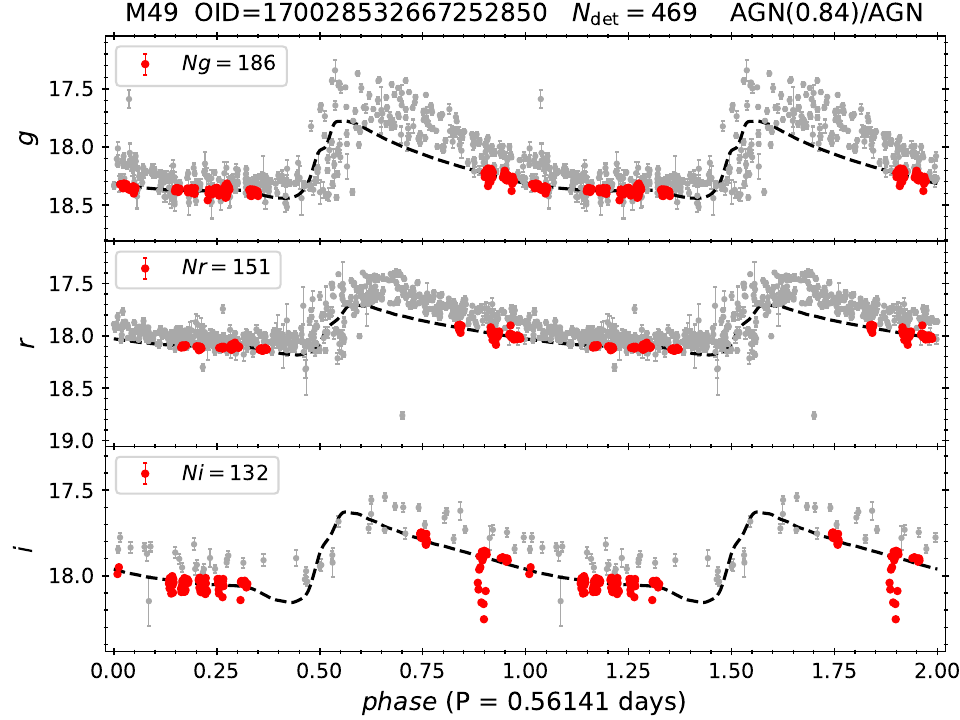}
  \caption{Comparison of the $gri$-band light curves extracted from Rubin alerts (red points) and the {\tt zubercal} online service (grey points). The light curves have been folded with the period adopted from \citet{sesar2013}, which exhibits the least scatter (as compared to the periods adopted from the PS1 RRL catalog, or other published periods). Based on the {\tt zubercal} light curves, this variable is most likely a Blazhko RRL.}
  \label{fig_2850}
\end{figure}

For the candidate associated with alert 170028532667252850, despite having a low $S_{PS1}$ score and being classified as AGN by the two alert brokers, it is a bona fide ab-type RRL. This is because it has been cataloged in various sky surveys and projects observed in different filters, including the Gaia DR3 RRL catalog and the work of \citet{drake2013}, \citet{sesar2013}, \citet{zinn2014}, \citet{chen2020}, \cite{chan2022} and \citet{huang2022}. The periods reported in these works are all consistent with a period of $\sim0.5614$~days. Its light curves extracted from the alerts miss the ascending branch and the descending branch after the maximum light, and hence at first glance the light curves did not appear to be an RRL. To confirm its RRL status, we compared the alert light curves and the $gri$-band light curves downloaded from the {\tt zubercal} online service,\footnote{\url{http://atua.caltech.edu/ZTF/Zubercal.html}} which were based on the data collected from the Zwicky Transient Facility \citep[ZTF,][]{bellm2019,graham2019}. The comparison of these two sets of light curves is presented in Figure \ref{fig_2850}, showing that this candidate is indeed an ab-type RRL.

\section{Discussions and Conclusions} \label{sec_last}

In this pilot study, we used Rubin early alerts to verify (or falsify) some of the RRL candidates that are fainter than $\sim 16$~mag in the public PS1, DES, and NGVS RRL catalogs. Of the 40 candidates identified in the alerts with at least 50 detections, eight of them are confirmed to be non-RRL, representing a fraction of 20\% of the sample.\footnote{If the two spectroscopically confirmed AGN were excluded, the fraction was reduced to $6/38\sim 0.16$ (or $\sim 16\%$).} This fraction is comparable to the findings of \citet{feng2026}, who found a fraction of $\sim 15\%$ from their sample. This implies that the public RRL catalogs, as well as the future RRL observations from Rubin-LSST, would have contamination at a similar level, especially toward the faint end, as the (expected) number of halo RRL would decrease at the same time as the number of AGN increases.   

Two alert brokers, {\tt ALeRCE} and {\tt Lasair}, have classified the alerts into broad categories, including VS and AGN. Together with a second classification scheme, which is based on scores from the PS1, DES, and NGVS RRL catalogs, these three classifications provide an initial assessment of whether a given alert or candidate is RRL or not. The final classifications, on the other hand, were based on visual inspection of the shapes of the multiband light curves extracted from the alerts; in some cases, we have also included the available spectral types. We found that candidates for which all three classifiers predicted ``VS'' label are confidently RRL. Otherwise, a candidate might not be a true RRL, and further information (e.g., light curve shape and/or spectral types) is needed to classify it. We also found an example in which all three classifiers failed to identify a genuine RRL.

Of the 40 alerts, the {\tt ALeRCE} stamp-based classifier correctly classified 28 (70\%) of them into the correct categories of VS (including binaries) and AGN. {\tt ALeRCE} classification is expected to improve once its light curve-based classifier \citep[e.g., see][]{ss2021} is activated on Rubin alerts. For {\tt Lasair}, the contextual based classifier correctly classified 17 (42.5\%) alerts, the rest are mainly classified as AGN even though they are RRL. Additional catalogs could be incorporated into the {\tt Sherlock} framework to improve classification. 

The non-detection of alerts on some of the candidates is interesting, which would imply those candidates are probably not RRL stars. This might also be true for alerts that only have one or two detections, as spurious detection might arise from imperfect image subtraction. In our pilot study, we only downloaded the alerts collected before 08 May 2026, and detections for them may increase in the near future which would allow a more detailed analysis.

\begin{acknowledgments}

  CCN acknowledges the funding from the National Science and Technology Council (NSTC, Taiwan) under Grant 114-2112-M-008-011. TARA is supported by the NSTC Grant 113-2740-M-008-005. AB acknowledges the funding from the Anusandhan National Research Foundation (ANRF) under the Prime Minister Early Career Research Grant scheme (ANRF/ECRG/2024/000675/PMS). This research was supported by the International Space Science Institute (ISSI) in Bern/Beijing through ISSI/ISSI-BJ International Team project ID $\#$24-603 – ``EXPANDING Universe'' (EXploiting Precision AstroNomical Distance INdicators in the Gaia Universe). We thank the useful discussions and comments from an anonymous referee to improve the manuscript.

  This research has made use of the SIMBAD database and the VizieR catalogue access tool, operated at CDS, Strasbourg, France. This research made use of Astropy,\footnote{\url{http://www.astropy.org}} a community-developed core Python package for Astronomy \citep{2013A&A...558A..33A,2018AJ....156..123A,2022ApJ...935..167A}.
 
  This material is based upon work supported in part by the National Science Foundation through Cooperative Agreements AST-1258333 and AST-2241526 and Cooperative Support Agreements AST-1202910 and 2211468 managed by the Association of Universities for Research in Astronomy (AURA), and the Department of Energy under Contract No. DE-AC02-76SF00515 with the SLAC National Accelerator Laboratory managed by Stanford University. Additional Rubin Observatory funding comes from private donations, grants to universities, and in-kind support from LSST-DA Institutional Members.

This publication is co-funded by
the European Union’s Horizon Europe research and innovation program under the Marie Sklodowska-Curie COFUND Postdoctoral Programme grant agreement No.101081355-SMASH and by the Republic of Slovenia and the European Union from the European Regional Development Fund. Views and opinions
expressed are however those of the author(s) only and do not necessarily reflect those of the European Union or European Research Exacutive Agency. Neither the European Union nor the granting authority can be held responsible for them.
  
\end{acknowledgments}

%\begin{contribution}

%\end{contribution}


\facilities{Rubin:Simonyi}


\software{{\tt ALeRCE} \citep{forster2021}, {\tt apply\_ugrizy\_templates.py} \citep{braga2024}, {\tt AstroInspect} \citep{cardoso2026}, {\tt Lasair} \citep{williams2024}, {\tt Sherlock} \citep{Young_sherlock_2023}, {\tt zubercal}}


\bibliography{VerifyRRL}{}
\bibliographystyle{aasjournal}



\end{document}